\documentclass[aps,prc,superscriptaddress,reprint]{revtex4-2}
\usepackage{graphicx}
\usepackage{multirow}
\usepackage{color}

\usepackage{amssymb}
\usepackage{bbm}
\usepackage{amsmath}
\usepackage{color}
\usepackage{lineno}
\usepackage{enumitem}
\usepackage{scalerel,stackengine}
\usepackage{xcolor}
\usepackage{comment}
\usepackage[normalem]{ulem}

\def\nuc#1#2{\relax\ifmmode{}^{#1}{\protect\text{#2}}\else${}^{#1}$#2\fi}

\newcommand{\be}{\begin{eqnarray}}
\newcommand{\ee}{\end{eqnarray}}

\newcommand{\bwt}{\begin{widetext}}
\newcommand{\ewt}{\end{widetext}}

\stackMath
\newcommand\reallywidehat[1]{%
\savestack{\tmpbox}{\stretchto{%
  \scaleto{%
    \scalerel*[\widthof{\ensuremath{#1}}]{\kern-.6pt\bigwedge\kern-.6pt}%
    {\rule[-\textheight/2]{1ex}{\textheight}}
  }{\textheight}%
}{0.5ex}}%
\stackon[1pt]{#1}{\tmpbox}%
}

\begin{document}


\title{Numerical Assessment of Convergence in the Post Form Ichimura-Austern-Vincent model}


\author{Jin Lei}
\email[]{jinl@tongji.edu.cn}

\affiliation{School of Physics Science and Engineering, Tongji University, Shanghai 200092, China.}




\begin{abstract}
The Ichimura-Austern-Vincent (IAV) model provides a powerful theoretical framework for describing inclusive breakup reactions. However, its post-form representation presents significant numerical challenges due to the absence of a natural cutoff in the transition matrix integration. This work presents a systematic assessment of convergence methods for post-form IAV calculations, comparing the bin method and the Vincent-Fortune approach. We demonstrate that while the bin method offers implementation simplicity, it exhibits strong parameter dependence that compromises numerical stability. In contrast, the Vincent-Fortune method, which employs complex contour integration, achieves reliable convergence without arbitrary parameters. We further introduce a novel hybrid approach that integrates finite-range distorted wave Born approximation (DWBA) with the Vincent-Fortune technique, combining the accuracy of finite-range treatment at short distances with the numerical stability of zero-range approximations in the asymptotic region. Numerical results for deuteron and $^6$Li-induced reactions confirm the efficacy of this hybrid method, showing consistent agreement with experimental data while eliminating the convergence issues that plague traditional approaches. This advancement enables more reliable calculations of nonelastic breakup cross sections and facilitates the extension of the IAV formalism beyond DWBA to incorporate continuum-discretized coupled-channels (CDCC) wave functions for a more comprehensive treatment of breakup processes.
\end{abstract}


\date{\today}%
\maketitle

\section{Introduction}\label{sec:intro}
In nuclear physics, the breakup process is a fundamental phenomenon wherein a projectile nucleus disintegrates into smaller fragments upon colliding with a target nucleus. This process is essential for probing nuclear structure, yielding critical insights into properties such as separation energies, angular momenta, parities, and electric responses to the continuum~\cite{BAUR1984333,TANIHATA1995505,Grinyer11,AUMANN2021103847,yang21,Li23}. Beyond structural exploration, breakup studies illuminate the reaction mechanisms governing weakly bound systems, facilitate the calculation of incomplete fusion cross sections, enable the determination of neutron-induced cross sections for short-lived nuclei through surrogate reactions, and enhance our understanding of nuclear astrophysics and the synthesis of heavy isotopes~\cite{PANDIT2021136570,Jutta12,Tumino21,TUMINO2025104164,Jin25}.

For two-body projectiles, the breakup reaction is typically denoted as $a + A \to b + x + A$, where the projectile $a$ comprises fragments $b$ and $x$. When all three outgoing particles, $b$, $x$, and the target $A$, are detected in a well-defined final state, the reaction is classified as \textit{exclusive}. Such reactions can be modeled as an effective three-body scattering problem using a suitable interaction potential. A rigorous theoretical approach to this problem is provided by the Faddeev equations~\cite{faddeev1965,THOMPSON200487,Deltuva13,Mukhamedzhanov12}, which can also incorporate excitations of the target or core. However, the computational complexity of solving these equations restricts their use to specific cases. As a result, more practical methods, such as the Continuum-Discretized Coupled-Channels (CDCC) method~\cite{Austern89,Austern87} and the Distorted Wave Born Approximation (DWBA)~\cite{BAUR1976293,CHATTERJEE2001476}, have gained widespread adoption due to their balance of accuracy and feasibility.

In contrast, \textit{inclusive} reactions occur when the final state is not fully specified. For example, in a reaction of the form $A(a, bX)$, only one constituent of the projectile, say $b$, is observed, while $X$ represents the unobserved particles, rendering the reaction inclusive with respect to those components. The simplest contribution to the inclusive cross section is elastic breakup (EBU), where all three outgoing particles remain in their ground states. However, more complex processes can also contribute, including breakup with excitation of $x$ or $A$, particle transfer between $x$ and $A$, or incomplete fusion (ICF), where $x$ fuses with $A$. These additional processes collectively constitute nonelastic breakup (NEB).

Given the multitude of possible final states, explicitly computing the NEB contribution by summing over all processes is generally impractical. To address this, several research groups in the 1980s developed closed-form expressions that leverage the completeness of the $x + A$ states to formally sum over final states~\cite{IAV85,UT81,HM85}. Among these, the model proposed by Ichimura, Austern, and Vincent (IAV) has attracted significant interest in recent years. Multiple groups have revisited and implemented the IAV model, demonstrating promising agreement with experimental data, a distinction not consistently achieved by competing approaches~\cite{Potel15,Torabi_2023,Jin15,Jin15b,Jin17,Jin18,Potel2017}.

The original IAV model employs closed-form expressions with post-form interactions. However, a notable limitation of this representation is the absence of a natural cutoff in the integration of the transition matrix, leading to non-converging numerical results due to long-range effects. To mitigate this, Huby and Mines introduced a convergence factor, $e^{-\alpha r}$~\cite{HUBY}, while Thompson proposed the bin method~\cite{Thompson_2011}. Both techniques, however, are highly sensitive to parameter choices, complicating efforts to achieve consistent convergence. Alternatively, Vincent and Fortune employed integration in the complex plane~\cite{Vincent-Fortune}, though this method is limited to the zero-range approximation with finite-range corrections. Consequently, practical applications of the IAV model often rely on its equivalent prior form. While the prior form performs adequately within the DWBA framework, it proves insufficient for capturing higher-order effects, such as those in Austern’s model using the three-body CDCC wave function, which I have recently implemented numerically with A.M. Moro~\cite{Jin19,Jin23}, necessitating a return to the post form.

In my previous work, I utilized the bin method to manage the post-form representation with both CDCC and DWBA wave functions~\cite{Jin15,Jin15b,Jin18,Jin19,Jin23,Jin25}. However, the results were sensitive to the choice of matching radius and bin size, and convergence was not consistently achieved across all parameter values. To address these convergence challenges, this paper proposes a detailed comparison between the bin method and the Vincent-Fortune method. Furthermore, I introduce a novel approach that integrates the finite-range DWBA with the Vincent-Fortune method, aiming to enhance convergence and improve the reliability of practical calculations. This study seeks to refine the computational framework for breakup reactions, advancing both theoretical understanding and experimental validation in nuclear physics.

The paper is organized as follows. In Section~\ref{sec:theory}, I present the theoretical framework of the IAV model, focusing on the formulation of nonelastic breakup reactions and the challenges associated with the post-form representation. Section~\ref{sec:method} introduces the convergence methods investigated in this work, including a detailed exposition of the novel hybrid approach that combines finite-range DWBA with the Vincent-Fortune technique. In Section~\ref{sec:results}, I present numerical applications to deuteron and $^6$Li-induced reactions, systematically comparing the performance of the bin method and the Vincent-Fortune method. Section~\ref{sec:discussion} offers a critical discussion of the strengths and limitations of each method, with particular emphasis on their applicability to more complex systems involving CDCC wave functions. Finally, Section~\ref{sec:conclusion} summarizes the findings and outlines directions for future research.

\section{Theoretical framework}\label{sec:theory}
The process under study involves a projectile labeled $a$, which has a two-body structure denoted as $a = b + x$. This projectile collides with a target nucleus $A$, leading to the emission of a fragment $b$. In this reaction, $b$ acts as a spectator, while $x$ is the participant interacting with $A$. The process can be represented as:
\begin{equation*}
a (= b + x) + A \to b + B^*,
\end{equation*}
where $B^*$ denotes any resulting state of the $x + A$ system.

Applying energy conservation in Jacobi coordinates, we obtain:
\begin{equation*}
E_{bx} + E_a = E_b + E_x,
\end{equation*}
where $E_{bx}$ is the relative energy of the $b + x$ pair, $E_x$ is the relative energy of the $x + A$ pair, $E_a$ is the relative energy of $a$ with respect to $A$, and $E_b$ is the relative energy of $b$ with respect to the $x + A$ system. The corresponding wave numbers for $E_a$, $E_b$, and $E_x$ are denoted $k_a$, $k_b$, and $k_x$, respectively.

The interaction between $x$ and $A$ encompasses both elastic scattering and nonelastic reactions. The elastic scattering is termed EBU, while the nonelastic processes, collectively referred to as NEB, include inelastic scattering of $x + A$, nucleon exchange between $x$ and $A$, fusion, and transfer to bound states of $B$.

In the three-body model proposed by Ichimura, Austern, and Vincent (IAV), the differential cross section for the NEB inclusive process is given by the closed-form expression:
\begin{equation}
\left. \frac{d^2\sigma}{dE_b \, d\Omega_b} \right|_{\text{NEB}} = -\frac{2}{\hbar v_a} \rho_b(E_b) \langle \varphi_x (\mathbf{k}_b) | \text{Im}[U_{xA}] | \varphi_x (\mathbf{k}_b) \rangle,
\end{equation}
where $\rho_b(E_b)$ is the density of states of particle $b$, $v_a$ is the velocity of the incoming projectile $a$, $\varphi_x(\mathbf{k}_b, \mathbf{r}_{xA})$ is the relative wave function describing the motion between $x$ and $A$ when $b$ is scattered with momentum $\mathbf{k}_b$, and $U_{xA}$ is the effective optical potential between $x$ and $A$. The wave function $\varphi_x(\mathbf{k}_b, \mathbf{r}_x)$ is determined by:
\begin{equation}
\label{eq:inh}
\varphi_x(\mathbf{k}_b, \mathbf{r}_x) = \int_0^{R_\textit{max}} G_x (\mathbf{r}_x, \mathbf{r}'_x) \langle \mathbf{r}'_x | \rho \rangle \, d\mathbf{r}'_x,
\end{equation}
where the source term is:
\begin{equation}
\label{eq:source}
\langle \mathbf{r}_x | \rho \rangle = \langle \mathbf{r}_x \chi_b^{(-)}(\mathbf{k}_b) | \mathcal{V}_{\text{post}} | \Psi^{3b(+)} \rangle.
\end{equation}
Here, $G_x$ is the Green's function incorporating the optical potential $U_{xA}$, $\chi_b^{(-)*}(\mathbf{k}_b, \mathbf{r}_b)$ is the distorted wave describing the relative motion of $b$ with respect to the $B^*$ system (obtained using an optical potential $U_{bB}$), and $\Psi^{3b(+)}$ is the exact three-body scattering wave function. The post-form transition operator is defined as $\mathcal{V}_{\text{post}} = V_{bx} + U_{bA} - U_{bB}$, where $V_{bx}$ is the binding potential of the $b + x$ projectile, and $U_{bA}$ is the optical potential for the relative scattering of $b$ and $A$. In the asymptotic limit, the wave function $\varphi_x(\mathbf{k}_b, \mathbf{r}_x)$ takes the form:
\begin{equation}
\varphi_x(\mathbf{k}_b, \mathbf{r}_x) \underset{r_x \to \infty}{\longrightarrow} f(\hat{k}_b, \hat{r}_x) \frac{e^{i k_x r_x}}{r_x},
\end{equation}
where $f(\hat{k}_b, \hat{r}_x)$ is the scattering amplitude, which can be used to compute the EBU cross section, as detailed in Ref.~\cite{Jin15}.

In the three-body model, the exact three-body wave function $\Psi^{3b(+)}$ is often approximated using the DWBA, expressed as $\Psi^{\text{DWBA}(+)} = \chi_a^{(+)} \phi_a$, where $\chi_a^{(+)}$ describes the elastic scattering of $a + A$, and $\phi_a$ is the bound-state wave function of the projectile $a$. Although the original IAV model employed the DWBA framework~\cite{IAV85}, Austern et al.~\cite{Austern87} extended it by incorporating the CDCC method, which includes continuum states of the $b + x$ system. For simplicity, this paper focuses on the DWBA form, which has been shown to be a reliable approximation of the CDCC wave function~\cite{Jin19,Jin23,Jin25}.

In configuration space, evaluating Eq.~(\ref{eq:inh}) poses numerical challenges, often leading to divergence. This issue arises because $V_{bx}$, the dominant component of $\mathcal{V}_{\text{post}}$, shares the same spatial coordinate as $\phi_a$, resulting in an unbounded $r_x$ integration without a natural cutoff. To address this, Huby and Mines~\cite{HUBY} and Vincent~\cite{Vincent68} introduced a convergence factor, redefining the source term as:
\begin{equation}
\langle \mathbf{r}_x | \rho \rangle \equiv \lim_{\alpha \to 0^+} \langle \mathbf{r}_x | e^{-\alpha \mathbf{r}_x} \rho \rangle.
\end{equation}
The damping factor $e^{-\alpha \mathbf{r}_x}$ ensures numerical stability. In principle, converged results are obtained by computing $\langle \mathbf{r}_x | e^{-\alpha \mathbf{r}_x} \rho \rangle$ for various $\alpha$ values and extrapolating to the limit $\alpha \to 0^+$. In practice, however, a small fixed $\alpha$ is typically chosen, making the numerical results sensitive to this parameter.

Another possible solution to this convergence problem is the bin method proposed by Thompson~\cite{Thompson_2011}, which rewrites the source term as:
\begin{equation}
\label{eq:bin}
\langle \mathbf{r}_x | \rho \rangle \approx \langle \chi_b^{(-)}(\mathbf{k}_b) | \chi_b^\mathrm{bin}(\bar{k}_b) \rangle \langle \mathbf{r}_x \chi_b^\mathrm{bin} (\bar{k}_b) | \mathcal{V}_{\text{post}} | \Psi^{3b(+)} \rangle,    
\end{equation}
with:
\begin{equation}
\langle \chi_b^\mathrm{bin}(\bar{k}_b) | \mathbf{r}_b \rangle = \sqrt{\frac{2}{\pi (k_{n+1} - k_n)}} \int_{k_n}^{k_{n+1}} \langle \chi_b^{(-)}(\mathbf{k}_b) | \mathbf{r}_b \rangle \, d\mathbf{k}_b.
\end{equation}
Since the bin wave function $\chi_b^\mathrm{bin}(\bar{k}_b)$ is square-integrable, it resolves the convergence problem in Eq.~(\ref{eq:inh}). However, the starting point of Eq.~(\ref{eq:bin}) involves inserting a complete basis of bin states and assuming this basis approximates the unit operator. With energy conservation, only the bin state covering the scattering state $\chi_b^{(-)}(\mathbf{k}_b)$ remains. Thus, the numerical results strongly depend on the bin size $\Delta k = k_{n+1} - k_n$. When the bin size is too small, a large integration interval is required to achieve convergence. Conversely, if the bin size is too large, the assumption that the basis of bin states equals the unit operator may not hold, potentially leading to misleading results. Therefore, achieving convergence requires extensive parameter tuning, balancing the matching radius and bin size.

Vincent and Fortune~\cite{Vincent-Fortune} introduce a distinct approach to transform the problematic real-axis integral into a more manageable form by employing contour integration in the complex radius plane. To achieve this, they divide the integration range into two regions: an inner region, where the integration can be evaluated numerically using standard methods, and an outer region, where the results are obtained through complex contour integration. Unlike the scaling factor method or the bin method, the Vincent-Fortune method is parameter-free, meaning its numerical results should not depend on arbitrarily chosen parameters. However, it requires special treatment of the outer region, where the scattering wave function is expressed in terms of outgoing ($H^+$) and incoming ($H^-$) Hankel functions. Consequently, the Vincent-Fortune method is currently limited to the zero-range approximation with finite-range corrections. Extending it to the finite-range form necessitates a coordinate transformation from incoming to outgoing Jacobi coordinates, which significantly complicates the handling of Hankel functions across different Jacobi coordinate systems.

\section{Convergence Method}\label{sec:method}
In this paper, I present a novel method for computing the relative wave function $\varphi_x(\mathbf{k}_b, \mathbf{r}_x)$ in Eq.~(\ref{eq:inh}), building upon the Vincent-Fortune approach. The reaction we are studying occurs in a regime where short-range nuclear forces dominate, typically acting over distances of just a few femtometers. These forces bind the projectile $a = b + x$, and the zero-range approximation simplifies this interaction by treating the potential $V_{bx}$ as a delta function, meaning $b$ and $x$ only interact when they are at the same point in space. While this assumption is a simplification, it becomes increasingly reasonable at large radial distances ($r_x \to \infty$), where the finite-range DWBA source term $\langle \mathbf{r}_x | \rho \rangle$, which captures the extended spatial structure of the $b + x$ system, approaches the simpler zero-range form. This behavior aligns with the asymptotic limit of the wave function, where $\varphi_x(\mathbf{k}_b, \mathbf{r}_x) \propto e^{i k_x r_x}/r_x$, reflecting the outgoing scattering of $x$ relative to $A$.

Inspired by this observation, the proposed method splits the computation of $\varphi_x(\mathbf{k}_b, \mathbf{r}_x)$ into two regions, leveraging the strengths of both finite-range and zero-range descriptions. Specifically, we express the wave function as:
\begin{align}
\label{eq:VF}
    \varphi_x(\mathbf{k}_b, \mathbf{r}_x) \approx & \int_0^R G_x (\mathbf{r}_x, \mathbf{r}'_x) \langle \mathbf{r}'_x | \rho^\mathrm{exact} \rangle \, d\mathbf{r}'_x \notag \\
    & + \int_R^\infty G_x (\mathbf{r}_x, \mathbf{r}'_x) \langle \mathbf{r}'_x | \rho^\mathrm{zero} \rangle \, d\mathbf{r}'_x,
\end{align}
where $G_x (\mathbf{r}_x, \mathbf{r}'_x)$ is the Green’s function incorporating the optical potential $U_{xA}$, $\langle \mathbf{r}'_x | \rho^\mathrm{exact} \rangle$ is the full finite-range DWBA source term, and $\langle \mathbf{r}'_x | \rho^\mathrm{zero} \rangle$ is its zero-range counterpart. The boundary $R$ separates the inner region (from 0 to $R$), where nuclear interactions are strong and the finite-range term is essential, from the outer region (from $R$ to $\infty$), where the zero-range approximation suffices due to the diminishing influence of $V_{bx}$. In the outer region, the integration is performed using the complex contour techniques of Vincent and Fortune, capitalizing on the asymptotic simplicity of the zero-range source term. The detailed partial wave decomposition of the source term can be found at Ref.~\cite{Jin18b}, and the partial wave decomposition of the zero-range source and the implementation of the Vincent-Fortune method can be found in the appendix.

This hybrid approach offers clear advantages. The use of the finite-range DWBA source term up to $R$ ensures accuracy in the inner region, where nuclear interactions shape the $x + A$ dynamics at short distances. Meanwhile, the zero-range source term enables efficient complex contour integration in the outer region, where long-range scattering dominates, avoiding the divergence issues of unbounded real-axis integrals. By integrating the parameter-free framework of the Vincent-Fortune method with a practical handling of finite-range effects, this new approach provides a robust and physically sound solution to the convergence challenges in Eq.~(\ref{eq:inh}). It ensures numerical stability without relying on ad hoc parameters like damping factors or bin sizes, ultimately enabling more precise calculations of breakup cross sections, such as those for elastic and nonelastic processes. This method thus represents a significant step forward in modeling three-body nuclear reactions with greater accuracy and reliability.

\section{Numerical Results}\label{sec:results}
In this section, I present calculations for reactions induced by deuteron and $^6$Li projectiles and compare the calculated NEB cross section with the post-form IAV model using the bin method and the Vincent-Fortune method. In all cases, I ignore the spin of the particles to simplify the calculations.

\subsection{Application to ($d$,$pX$)}
As a first example, I consider the reaction $^{62}$Ni($d$, $pX$) at an incident deuteron energy $E_d = 25.5$ MeV, which was previously analyzed in Ref.~\cite{Jin15b}. 

In the calculations, the deuteron ground-state wave function was generated using a simple Gaussian potential, as described in Ref.~\cite{Austern87}. The deuteron distorted waves were computed with an optical potential sourced from Ref.~\cite{yyh06}. For the proton-target and neutron-target interactions, I employed the global parametrization of Koning and Delaroche (KD) from Ref.~\cite{KD02}.

\begin{figure}[tb]
\begin{center}
 {\centering \resizebox*{0.9\columnwidth}{!}{\includegraphics{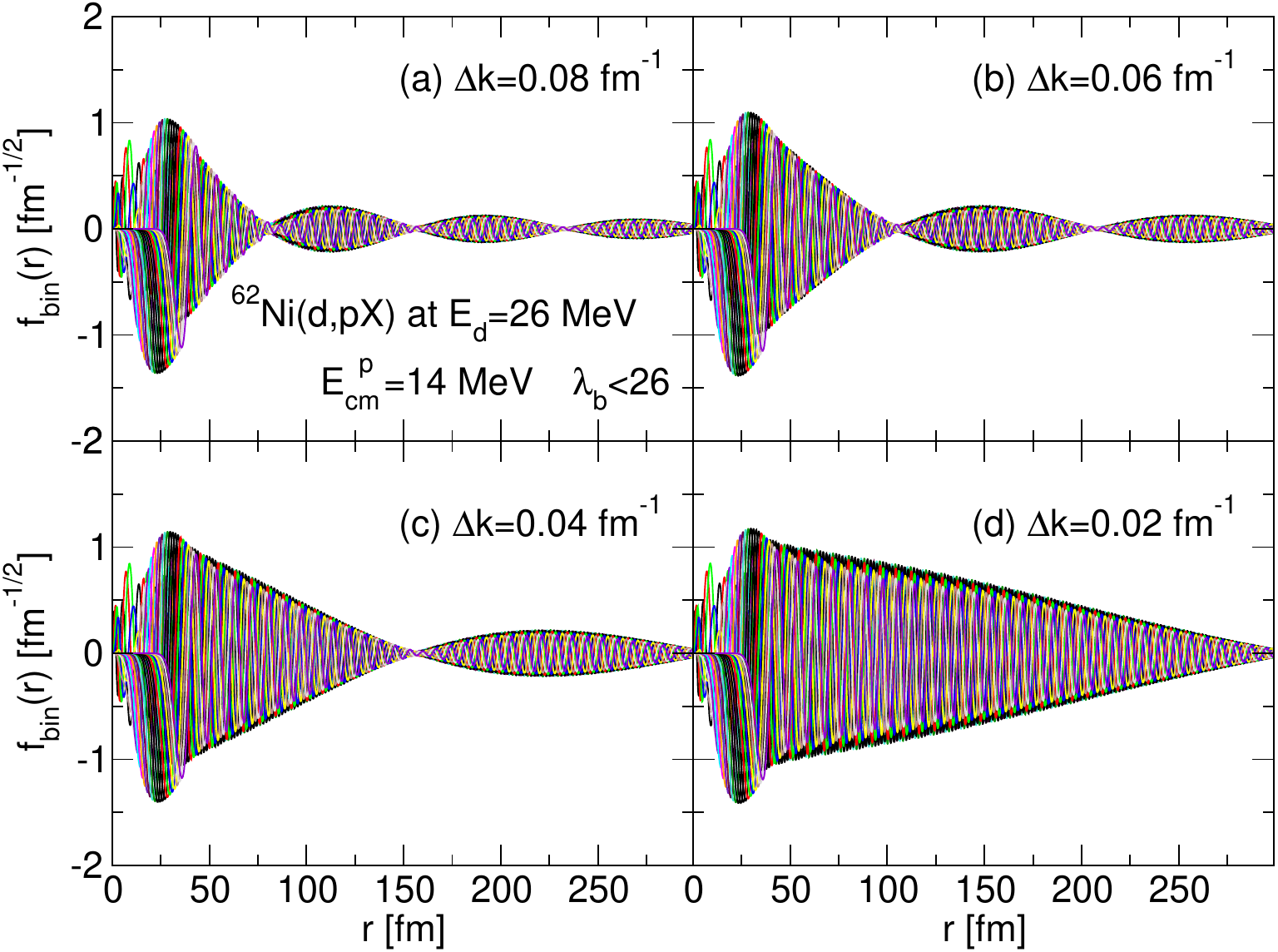}} \par}
\caption{\label{fig:compare_bin_wf}Radial parts of the bin state wave function at a proton energy of 14 MeV in the lab frame, calculated with different bin sizes: (b) $\Delta k = 0.06$ fm$^{-1}$; (c) $\Delta k = 0.04$ fm$^{-1}$; (d) $\Delta k = 0.02$ fm$^{-1}$.}
\end{center}
\end{figure}

As outlined in the previous section, evaluating the post-form formulas requires specialized techniques. Here, I first discuss the bin method, which involves averaging the distorted waves $\chi_b^{(-)}(\mathbf{k}_b, \mathbf{r}_b)$ over small momentum intervals (bins). This averaging renders the resulting functions square-integrable, making the source term in Eq.~(\ref{eq:bin}) short-ranged. Before presenting the NEB results from the post-form IAV model, I examine the radial part of the bin state wave function at a proton energy of 14 MeV in the lab frame, as shown in Fig.~\ref{fig:compare_bin_wf}. The figure displays the radial components for partial waves with $0 \leq \lambda_b \leq 25$, calculated using bin sizes of $\Delta k = 0.08$ fm$^{-1}$, 0.06 fm$^{-1}$, 0.04 fm$^{-1}$, and 0.02 fm$^{-1}$, labeled as (a), (b), (c), and (d), respectively. 

The figure reveals that as the bin size ($\Delta k$) increases, the bin state wave function becomes more spatially compressed, localizing within a smaller region. For example, at $\Delta k = 0.08$ fm$^{-1}$, individual wave packets are distinguishable, with the amplitudes of subsequent packets decreasing gradually until a square-integrable wave function emerges. This behavior aligns with the uncertainty principle: a larger $\Delta k$ corresponds to a smaller uncertainty in position, $\Delta r$. Conversely, at $\Delta k = 0.02$ fm$^{-1}$, the wave packet exhibits significant spatial broadening, resulting in a larger spread in position space.

\begin{figure}[tb]
\begin{center}
 {\centering \resizebox*{0.9\columnwidth}{!}{\includegraphics{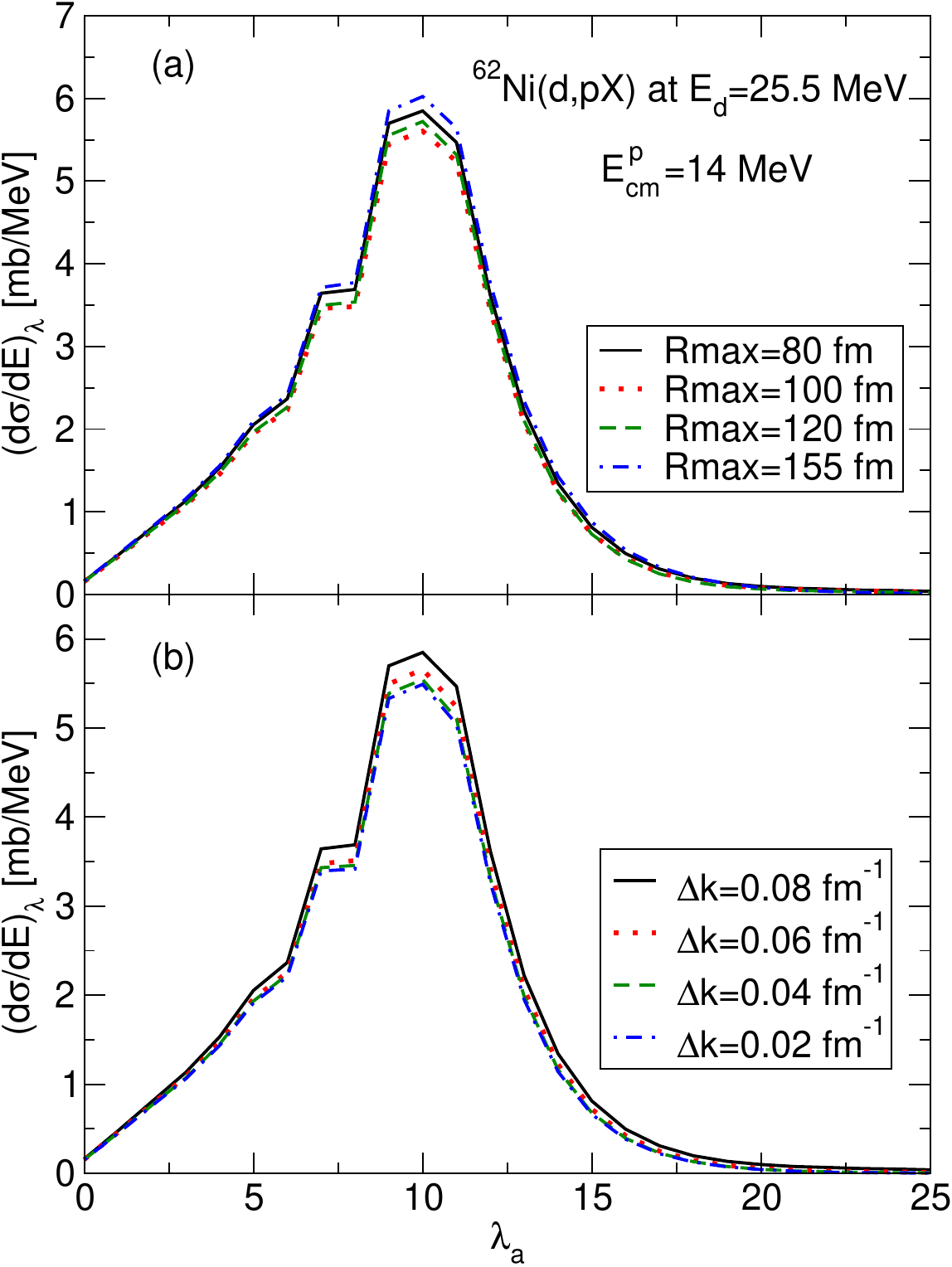}} \par}
\caption{\label{fig:com_wbin_R}Differential cross section energy distribution of the partial wave components of the NEB for the reaction $^{62}$Ni($d$, $pX$) at $E_d = 25.5$ MeV, with the outgoing proton energy at 14 MeV. (a) Fixed $\Delta k = 0.08$ fm$^{-1}$ with varying $R_{max}$ values, where $R_{max}$ is the maximum radial distance in the numerical calculations; (b) results with different $\Delta k$ values, each with $R_{max}$ set at the nodes of the outgoing proton wave functions. See text for details.}
\end{center}
\end{figure}

In principle, achieving fully converged results with the bin method requires setting the maximum integration range, $R_{\text{max}}$, to a value where the radial part of the bin state wave function effectively vanishes. However, this ideal scenario presents significant numerical challenges. For instance, as shown in Fig.~\ref{fig:compare_bin_wf}, even with the largest bin size considered ($\Delta k = 0.08$ fm$^{-1}$), an $R_{\text{max}}$ of 300 fm is insufficient to ensure the wave function fully diminishes.

To investigate this further, I analyze the differential cross section energy distribution for the partial wave components of the NEB in the reaction $^{62}$Ni($d$, $pX$) at $E_d = 25.5$ MeV, with the outgoing proton energy fixed at 14 MeV in the center-of-mass frame. These results are presented in Fig.~\ref{fig:com_wbin_R}(a), where calculations are performed with a fixed bin size of $\Delta k = 0.08$ fm$^{-1}$ and varying $R_{\text{max}}$ values: 80 fm (solid line), 100 fm (dotted line), 120 fm (dashed line), and 155 fm (dash-dotted line). For the partial wave $\lambda_a = 10$, significant discrepancies appear across these $R_{\text{max}}$ values, indicating that simply increasing $R_{\text{max}}$ does not guarantee convergence.

In practice, the bin method often sets $R_{\text{max}}$ at the nodes of the bin state wave function to confine it within a wave packet. For $\Delta k = 0.08$ fm$^{-1}$, these nodes occur at approximately 80 fm, 155 fm, 235 fm, and beyond. However, the results show a marked difference between $R_{\text{max}} = 80$ fm and $R_{\text{max}} = 155$ fm, highlighting numerical instability in the bin method. This suggests that relying solely on node-based $R_{\text{max}}$ selections does not consistently yield stable, converged results.

Figure~\ref{fig:com_wbin_R}(b) presents results for different $\Delta k$ values, with $R_{\text{max}}$ set to the first node of the outgoing proton wave function for each case: $\Delta k = 0.08$ fm$^{-1}$ with $R_{\text{max}} = 80$ fm (solid line), $\Delta k = 0.06$ fm$^{-1}$ with $R_{\text{max}} = 105$ fm (dotted line), $\Delta k = 0.04$ fm$^{-1}$ with $R_{\text{max}} = 155$ fm (dashed line), and $\Delta k = 0.02$ fm$^{-1}$ with $R_{\text{max}} = 310$ fm (dash-dotted line). Noticeable differences are evident among the results for $\Delta k = 0.08$ fm$^{-1}$, 0.06 fm$^{-1}$, and 0.04 fm$^{-1}$. Although larger $\Delta k$ values compress the wave function into a smaller spatial region, these bin sizes appear too coarse to accurately capture the features of the original scattering wave function. However, the results for $\Delta k = 0.04$ fm$^{-1}$ and $\Delta k = 0.02$ fm$^{-1}$ show negligible differences, suggesting convergence, consistent with findings in prior studies (e.g., Refs.~\cite{Jin15, Jin15b, Jin18, Jin19, Jin23, Jin25}).

\begin{figure}[tb]
\begin{center}
 {\centering \resizebox*{0.9\columnwidth}{!}{\includegraphics{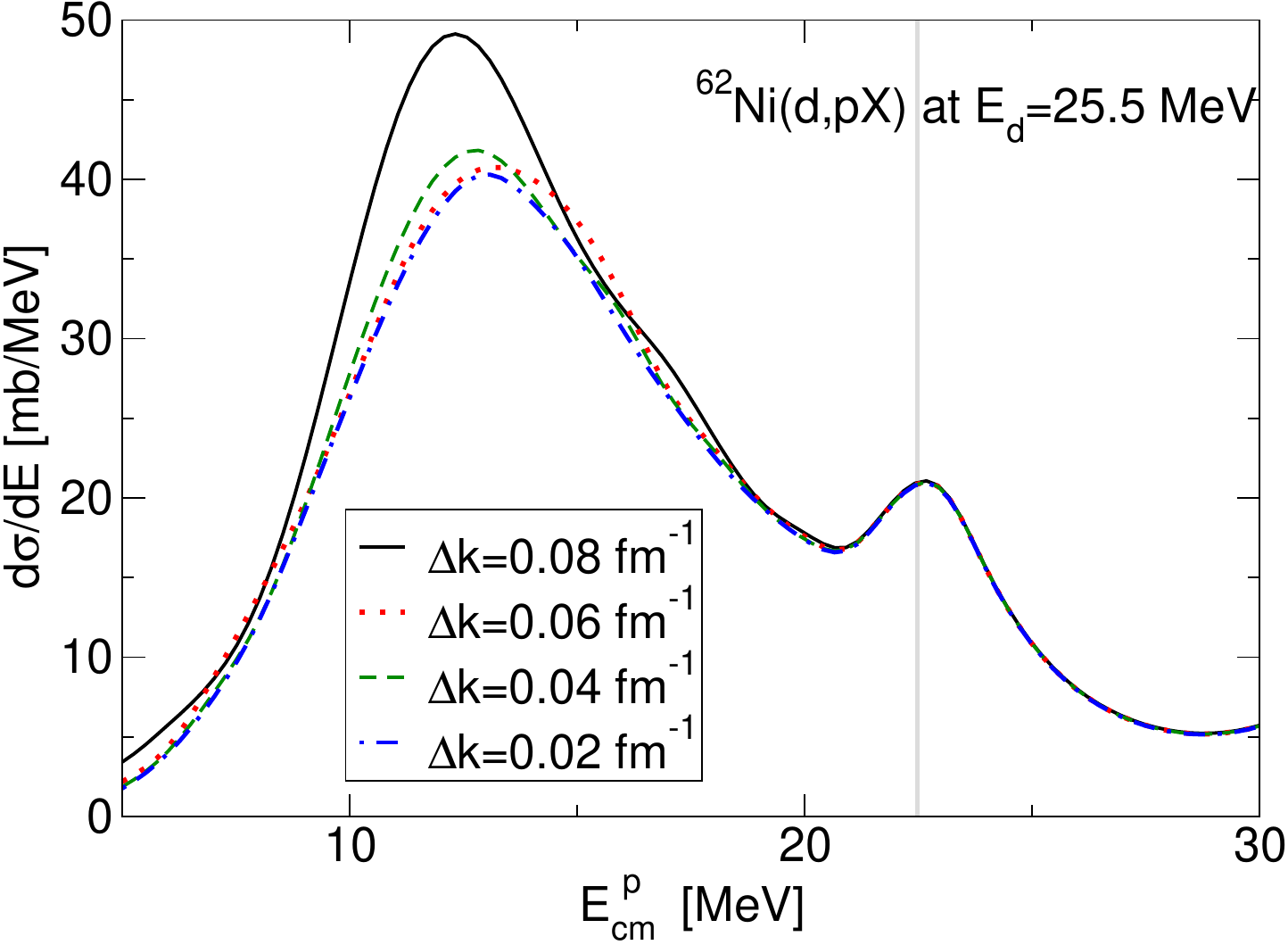}} \par}
\caption{\label{fig:com_dsde_bin}Differential cross section energy distribution of the NEB for the reaction $^{62}$Ni($d$, $pX$) at $E_d = 25.5$ MeV as a function of the outgoing proton energy. Results are obtained using different $\Delta k$ values, with $R_{\text{max}}$ set at the nodes of the outgoing proton wave functions for each $\Delta k$. See text for details.}
\end{center}
\end{figure}

Next, I present the differential cross section energy distribution of the NEB as a function of the outgoing proton energy in the center-of-mass frame, shown in Fig.~\ref{fig:com_dsde_bin}. The results correspond to $\Delta k = 0.08$ fm$^{-1}$ with $R_{\text{max}} = 80$ fm (solid line), $\Delta k = 0.06$ fm$^{-1}$ with $R_{\text{max}} = 105$ fm (dotted line), $\Delta k = 0.04$ fm$^{-1}$ with $R_{\text{max}} = 155$ fm (dashed line), and $\Delta k = 0.02$ fm$^{-1}$ with $R_{\text{max}} = 310$ fm (dash-dotted line). A vertical line marks the threshold where the relative energy between the neutron and $^{62}$Ni is zero. The differences between these calculations are more pronounced than those in Fig.~\ref{fig:com_wbin_R}(b). This arises because the positions of the first nodes of the scattering wave function in coordinate space shift as the outgoing proton energy changes. While the calculations in Fig.~\ref{fig:com_dsde_bin} use a fixed $R_{\text{max}}$ for simplicity, this approach proves inadequate for certain $\Delta k$ values in this context.

\begin{figure}[tb]
\begin{center}
 {\centering \resizebox*{0.9\columnwidth}{!}{\includegraphics{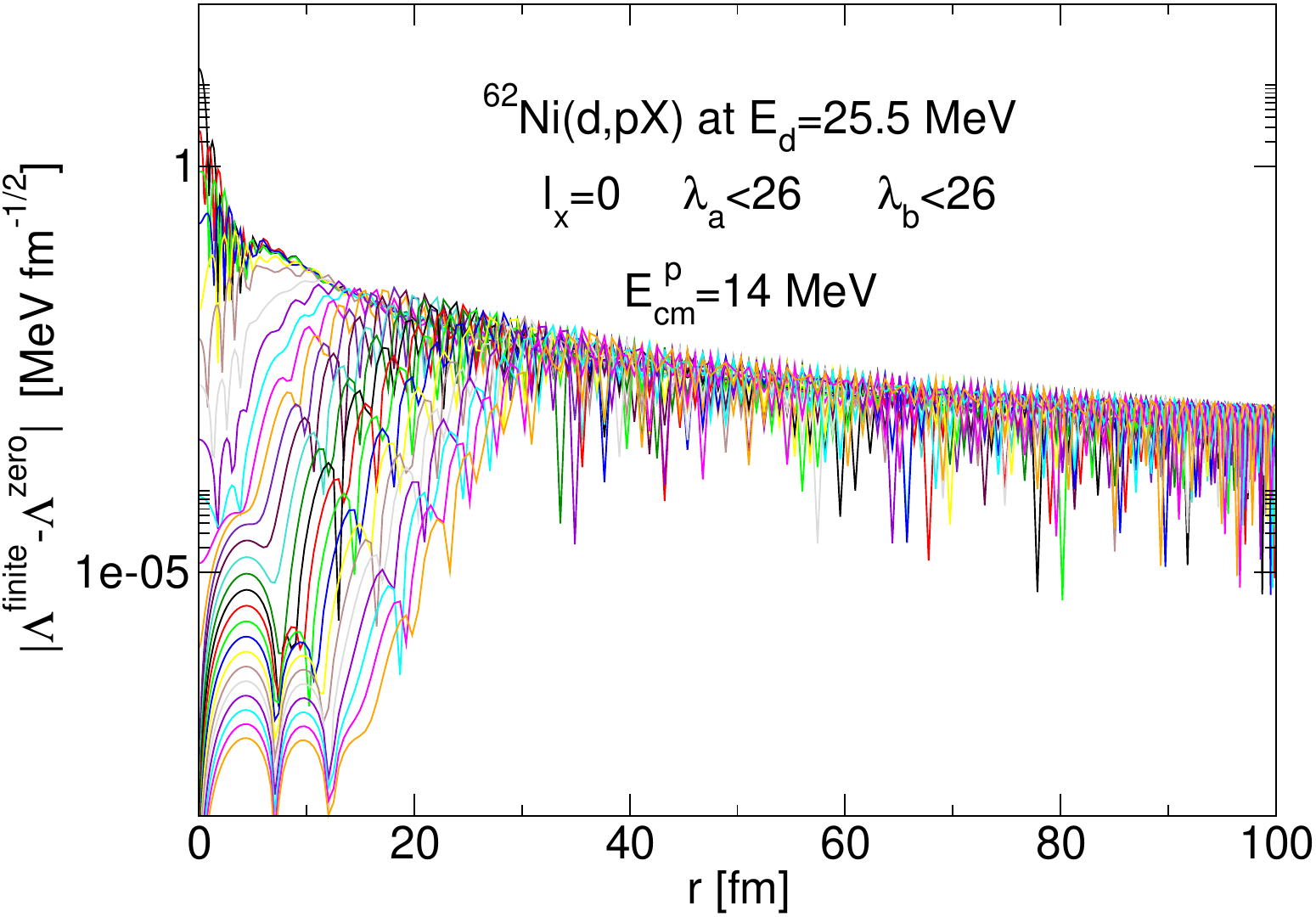}} \par}
\caption{\label{fig:com_source}Absolute value of the difference of the $\Lambda$ function, cf. Eq.~(\ref{eq:lamba}), computed using finite-range and zero-range DWBA for the reaction $^{62}$Ni($d$, $pX$) at $E_d = 25.5$ MeV, with an outgoing proton energy of 14 MeV in the center-of-mass frame, for partial waves with $l_x = 0$, $0 \leq \lambda_a \leq 25$, and $0 \leq \lambda_b \leq 25$.}
\end{center}
\end{figure}

I now turn to the Vincent-Fortune method, which combines finite-range and zero-range treatments of the source term function. The underlying assumption is that beyond a certain radial distance $R$, the finite-range and zero-range approximations yield equivalent results. To test this, Fig.~\ref{fig:com_source} shows the absolute difference of the $\Lambda$ function (the radial part of the source term, cf. Eq.~(\ref{eq:lamba})), calculated using finite-range and zero-range DWBA methods for partial waves with $l_x = 0$, $0 \leq \lambda_a \leq 25$, and $0 \leq \lambda_b \leq 25$. The difference decreases exponentially with increasing distance, supporting the validity of using the zero-range approximation in Eq.~(\ref{eq:VF}).

\begin{figure}[tb]
\begin{center}
 {\centering \resizebox*{0.9\columnwidth}{!}{\includegraphics{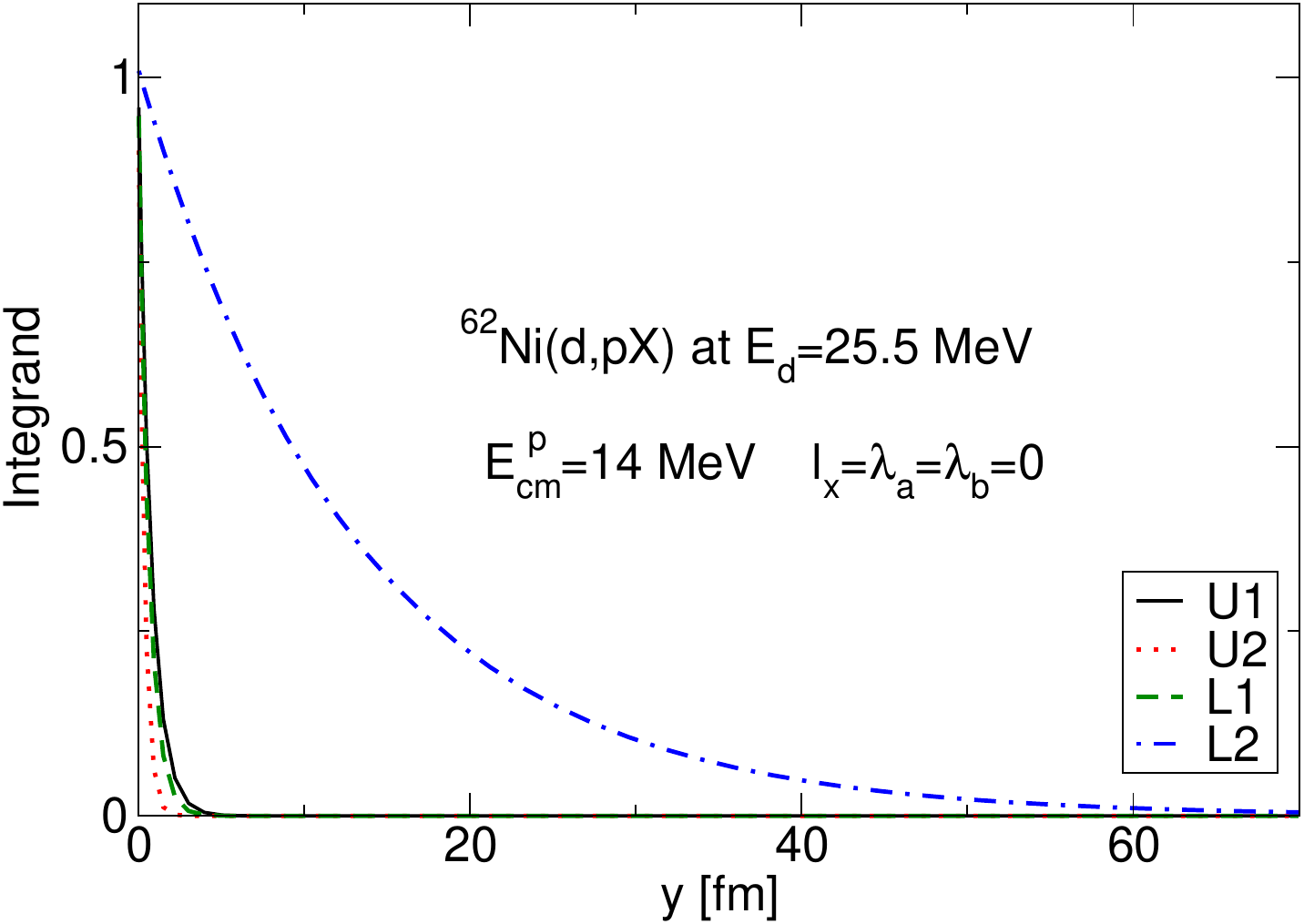}} \par}
\caption{\label{fig:com_imag_VF}Integrand function, fixed at $x = R$ on the complex plane, varying along the $y$-axis for the reaction $^{62}$Ni($d$, $pX$) at $E_d = 25.5$ MeV, with an outgoing proton energy of 14 MeV, for partial waves with $l_x = \lambda_a = \lambda_b = 0$. See text for details.}
\end{center}
\end{figure}

To assess the numerical stability of the Vincent-Fortune method, I plot the integrand functions from Eq.~(\ref{eq:integrand}) in Fig.~\ref{fig:com_imag_VF} for the same reaction, using partial waves with $l_x = \lambda_a = \lambda_b = 0$. The figure comprises two panels. In the upper panel, the solid line represents $U1 = |H_{l_x}^{(+)}(k_x z) H_{\lambda_b}^{(-)}(k_b c z) H_{\lambda_a}^{(+)}(k_a z)|$, and the dotted line shows $U2 = |H_{l_x}^{(+)}(k_x z) H_{\lambda_b}^{(+)}(k_b c z) H_{\lambda_a}^{(+)}(k_a z)|$. In the lower panel, the dashed line depicts $L1 = |H_{l_x}^{(+)}(k_x z) H_{\lambda_b}^{(-)}(k_b c z) H_{\lambda_a}^{(-)}(k_a z)|$, and the dash-dotted line represents $L2 = |H_{l_x}^{(+)}(k_x z) H_{\lambda_b}^{(+)}(k_b c z) H_{\lambda_a}^{(-)}(k_a z)|$. All four components ($U1$, $U2$, $L1$, and $L2$) are smooth functions. Three of them ($U1$, $U2$, and $L1$) are short-ranged, with significant contributions only up to approximately 7 fm. In contrast, $L2$ is long-ranged, extending along the $y$-direction to about 70 fm. This behavior stems from the exponential growth of $H_{l_x}^{(+)}(k_x z)$ and $H_{\lambda_b}^{(+)}(k_b c z)$ in the lower panel, partially offset by the faster exponential decay of $H_{\lambda_a}^{(-)}(k_a z)$. However, this decay does not fully compensate for the growth, slowing the convergence of $L2$ and allowing it to extend over a larger distance.

\begin{figure}[tb]
\begin{center}
 {\centering \resizebox*{0.9\columnwidth}{!}{\includegraphics{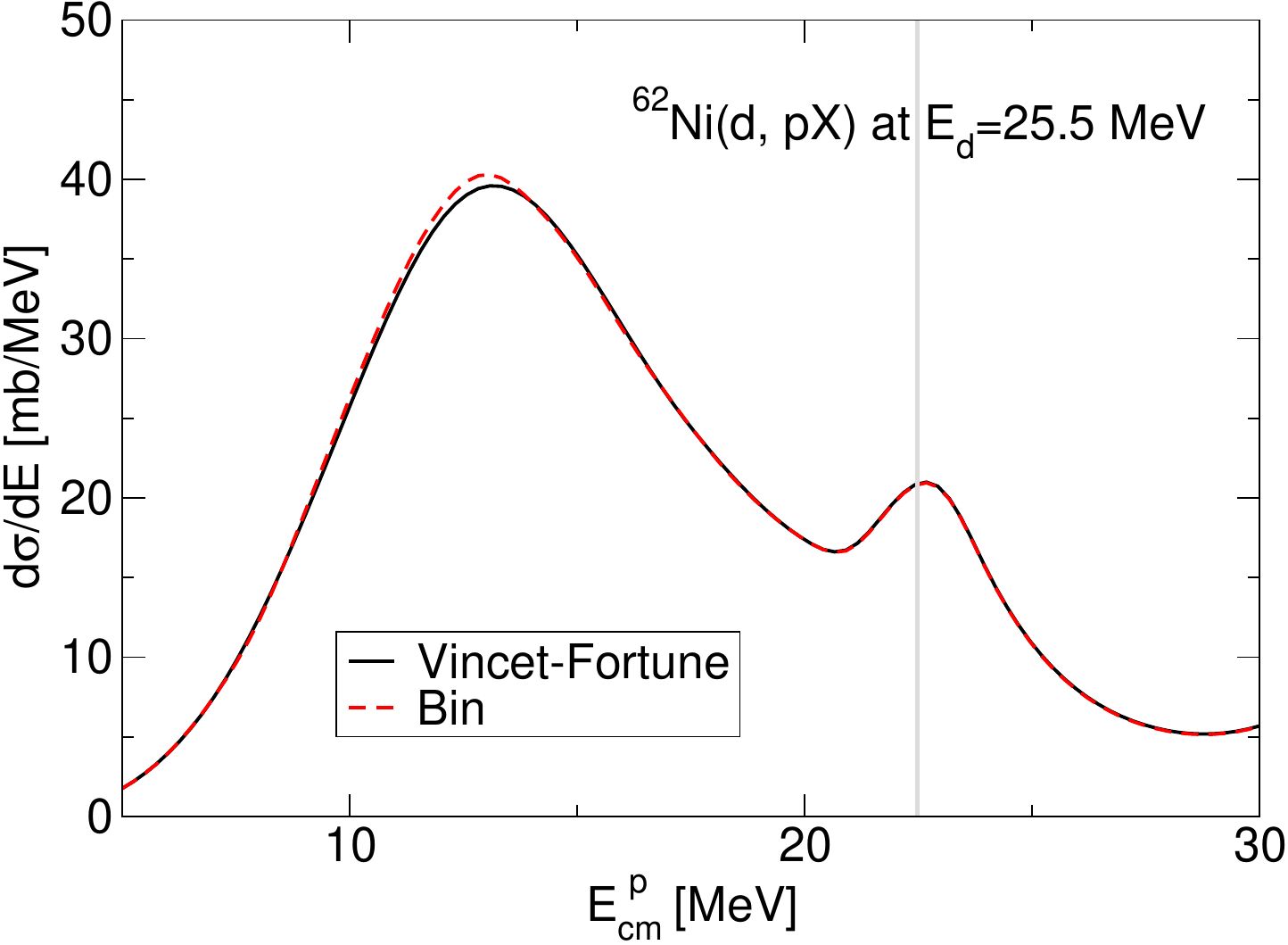}} \par}
\caption{\label{fig:com_VF_bin}Differential cross section energy distribution of the NEB for the reaction $^{62}$Ni($d$, $pX$) at $E_d = 25.5$ MeV as a function of the outgoing proton energy. Results are obtained using the bin method and the Vincent-Fortune method.}
\end{center}
\end{figure}

Finally, I compare the results from the bin method and the Vincent-Fortune method for the same reaction in Fig.~\ref{fig:com_VF_bin}. The solid line represents the bin method with $\Delta k = 0.02$ fm$^{-1}$, while the dashed line corresponds to the Vincent-Fortune method. The two sets of results show reasonable agreement. However, unlike the bin method, which depends on parameters like $\Delta k$ and $R_{\text{max}}$, the Vincent-Fortune method is parameter-independent and exhibits better numerical stability and convergence.

\subsection{Application to ($^6$Li,$\alpha X$)}
As a second example, I examine the reaction $^{209}$Bi($^6$Li, $\alpha X$), which was previously analyzed in Ref.~\cite{Jin15} using the post-form IAV model with the bin method. Those calculations successfully reproduced the experimental angular distribution of $\alpha$ particles across a wide range of incident energies, both above and below the Coulomb barrier. To assess the numerical stability of the bin method and compare its results with those obtained using the Vincent-Fortune approach, I consider an incident energy of $E = 50$ MeV. For the calculations presented here, I adopt the potentials used in Ref.~\cite{Jin15}.

\begin{figure}[tb]
\begin{center}
 {\centering \resizebox*{0.9\columnwidth}{!}{\includegraphics{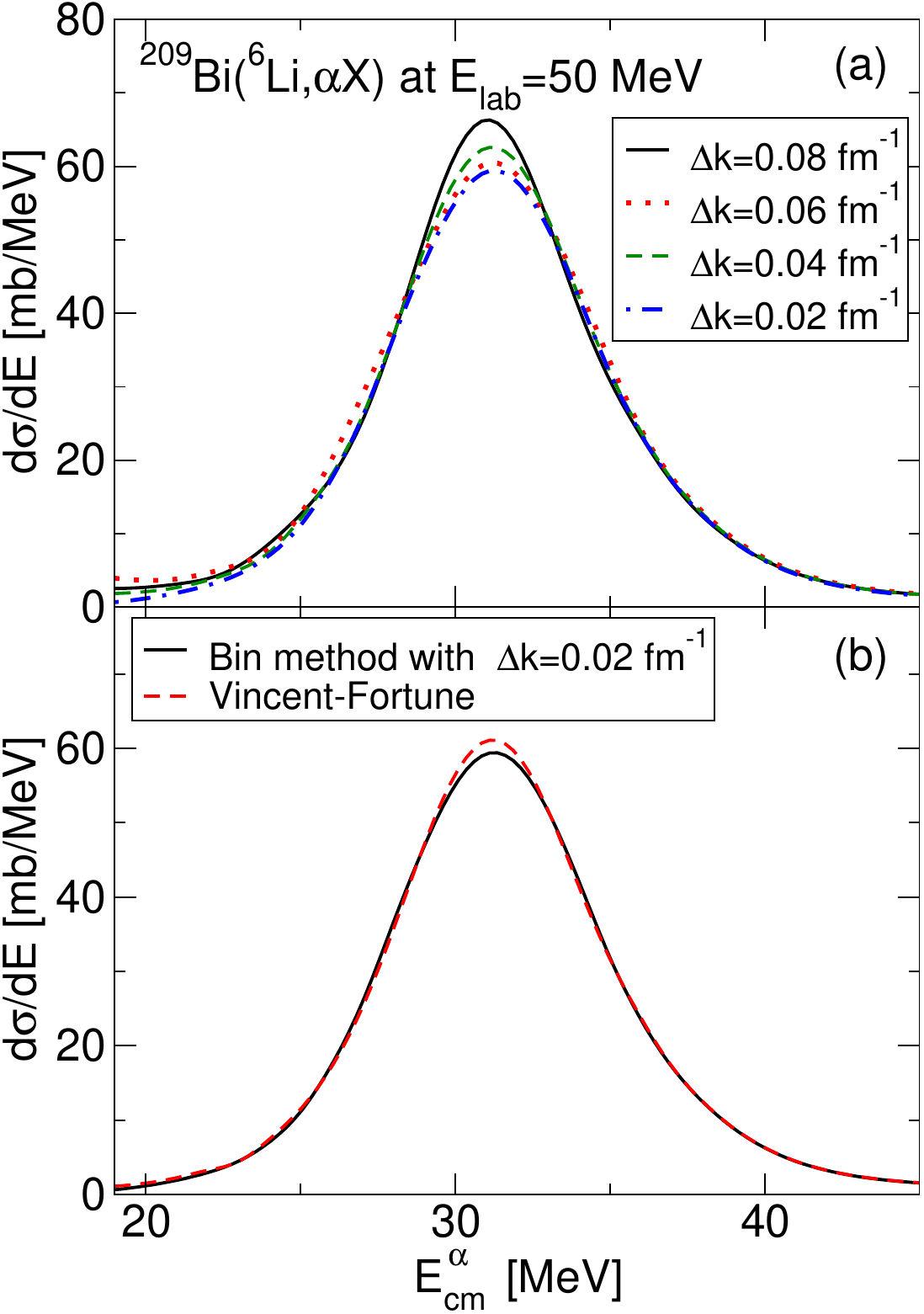}} \par}
\caption{\label{fig:com_VF_bin_6li}Differential cross section energy distribution of the NEB for the reaction $^{209}$Bi($^6$Li, $\alpha X$) at $E_{lab} = 50$ MeV as a function of the outgoing proton energy. Results are obtained using the bin method and the Vincent-Fortune method.}
\end{center}
\end{figure}

I first compare the differential cross-section energy distributions computed using the bin method with varying $\Delta k$ values, as shown in Fig.\ref{fig:com_VF_bin_6li} (a), for the reaction $^{209}$Bi($^6$Li, $\alpha X$) at $E_{\text{lab}} = 50$ MeV. The solid, dotted, dashed, and dash-dotted lines represent results with $\Delta k = 0.08$ fm$^{-1}$ and $R_{\text{max}} = 80$ fm, $\Delta k = 0.06$ fm$^{-1}$ and $R_{\text{max}} = 110$ fm, $\Delta k = 0.04$ fm$^{-1}$ and $R_{\text{max}} = 155$ fm, and $\Delta k = 0.02$ fm$^{-1}$ and $R_{\text{max}} = 310$ fm, respectively. Unlike the deuteron case, a clear trend toward convergence emerges as $\Delta k$ decreases, suggesting that the bin method is numerically unreliable. In Fig.\ref{fig:com_VF_bin_6li} (b), I present a comparison for the same reaction using the Vincent-Fortune method. The solid and dashed lines correspond to results from the bin method with $\Delta k = 0.02$ fm$^{-1}$ and $R_{\text{max}} = 310$ fm and the Vincent-Fortune method, respectively. In contrast to the bin method, the Vincent-Fortune method exhibits significantly greater numerical stability.

\section{Discussion}\label{sec:discussion}
As discussed above, the bin method, which averages distorted waves over momentum bins, is simple, flexible, and straightforward to implement by replacing the standard scattering wave function with a bin-state wave function. However, it is numerically unstable, exhibiting sensitivity to parameters such as $R_{\text{max}}$ (the maximum radial cutoff) and $\Delta k$ (the momentum bin size). This sensitivity leads to inconsistent results and poor convergence unless impractically small bin sizes and large radial ranges are employed. In contrast, the Vincent-Fortune method, which combines finite- and zero-range treatments, ensures rapid convergence and robustness without relying on arbitrary parameters. However, it assumes that the source term remains consistent beyond a radial distance $R$, an assumption that may not hold universally and is strictly valid only for $s$-wave bound-state projectiles. Additionally, its reliance on complex-plane integration can be computationally intensive. In practice, however, calculations using the prior-form IAV model can readily achieve convergence without numerical difficulties.

Challenges emerge when extending beyond the DWBA, such as when incorporating CDCC wave functions into the IAV model, as explored in Refs.~\cite{Jin19, Jin23, Jin25}. In these cases, studying convergence in the post-form formalism becomes critical. On the other hand, the zero-range approximation can also be applied to CDCC wave functions for two reasons. First, reaction systems requiring CDCC wave functions typically involve weakly bound projectiles with $s$-wave ground states, justifying the use of a zero-range source term. Second, for the full CDCC wave function, $\Psi^\mathrm{CDCC(+)}(\mathbf{r}_a) \xrightarrow{\mathbf{r}_a \to \infty} \chi_a(\mathbf{r}_a)$, which describes elastic scattering between the projectile $a$ and target $A$. The relative coordinate $\mathbf{r}_{bx}$ is naturally truncated due to the short-range nature of the $V_{bx}$ interaction, causing the source term computed with CDCC to approach results obtained from the zero-range approximation.

Future work should focus on testing the Vincent-Fortune method in complex systems (e.g., exotic nuclei or higher-energy reactions), optimizing its computational efficiency, and integrating it with CDCC wave functions. The Vincent-Fortune method represents a reliable, parameter-independent improvement over the bin method, offering significant potential for precision studies of breakup reactions when combined with CDCC wave functions.

\section{Summary and conclusions}\label{sec:conclusion}
In this paper, I have presented a comprehensive investigation of numerical convergence methods for the post-form IAV model, focusing on the calculation of nonelastic breakup reactions. The primary challenge addressed is the numerical instability inherent in the post-form representation, which lacks a natural cutoff in the integration of the transition matrix. Two main approaches were compared: the bin method and the Vincent-Fortune method, along with a novel hybrid approach that combines the finite-range DWBA with the Vincent-Fortune technique.

The bin method, which averages distorted waves over momentum bins to render them square-integrable, offers a straightforward implementation but exhibits significant parameter sensitivity. As demonstrated in calculations for the $^{62}$Ni($d$, $pX$) reaction at $E_d = 25.5$ MeV, the results depend critically on both the bin size ($\Delta k$) and the maximum integration radius ($R_{\text{max}}$). While convergence can be achieved with sufficiently small bin sizes (around $\Delta k = 0.02$ fm$^{-1}$), this requires impractically large integration radii (over 300 fm), introducing computational inefficiencies. The method's reliance on setting $R_{\text{max}}$ at the nodes of the bin state wave function further complicates its application, as these nodes shift with changing outgoing particle energies.

In contrast, the Vincent-Fortune method, which employs complex contour integration, demonstrates superior numerical stability without dependence on arbitrary parameters. The key insight underlying this approach is that beyond a certain radial distance, the finite-range and zero-range treatments of the source term yield equivalent results, as confirmed by the exponential decrease in their difference with increasing radius. The method divides the calculation into an inner region, where finite-range effects are essential, and an outer region, where zero-range approximations suffice, leveraging the asymptotic behavior of the wave function.

The proposed hybrid approach combines the strengths of both finite-range and zero-range descriptions, offering a robust solution to the convergence challenges in the post-form IAV model. For the reactions studied, $^{62}$Ni($d$, $pX$) at $E_d = 25.5$ MeV and $^{209}$Bi($^6$Li, $\alpha X$) at $E = 50$ MeV, this approach produces results consistent with the bin method at small bin sizes, but with greater numerical stability and computational efficiency.

Despite these advances, several challenges remain. The Vincent-Fortune method assumes that the source term remains consistent beyond a certain radial distance, an assumption that may not hold universally and is strictly valid only for $s$-wave bound-state projectiles. Additionally, complex-plane integration can be computationally intensive, though this is offset by the method's rapid convergence. For practical applications, the prior-form IAV model often remains a viable alternative, as it naturally achieves convergence without numerical difficulties.

Looking forward, the convergence techniques developed here hold particular significance for extending the IAV model beyond the DWBA framework. The integration of the Vincent-Fortune method with CDCC wave functions, as explored in recent work \cite{Jin19, Jin23, Jin25}, represents a promising direction for future research. This is especially relevant for systems involving weakly bound projectiles with $s$-wave ground states, where the zero-range approximation can be justified even for CDCC wave functions.

In conclusion, the hybrid approach combining finite-range DWBA with the Vincent-Fortune method represents a significant advancement in modeling three-body nuclear reactions. By providing a parameter-independent, physically motivated solution to the convergence challenges in the post-form IAV model, this method enhances the precision and reliability of breakup cross-section calculations, with broad implications for nuclear reaction studies.

\begin{acknowledgments}
I am grateful to Antonio Moro for his invaluable discussions. This work has been partially supported by the National Natural Science Foundation of China (Grants No. 12475132 and No. 12105204) and by the Fundamental Research Funds for the Central Universities.
\end{acknowledgments}

\bibliography{inclusive.bib}
\appendix
\onecolumngrid
\section{\label{sec:appendix}Partial Wave Analysis of the Zero-Range Source Term and Vincent-Fortune Method}

Instead of employing a three-dimensional Jacobi basis, the wave function can be expanded into partial wave eigenstates, which depend on the radial magnitude and angular momentum eigenstates. The orbital angular momenta of the three particles are coupled to a total angular momentum $J$ and its third component $M_J$. For the incoming channels, the state is expressed as
\begin{align}
\left| r_{bx} r_a \alpha_{\text{in}} M_J \right\rangle = \left| r_{bx} r_a \left( \left( l_a \left( j_b j_x \right) s_{bx} \right) J_a \left( \lambda_a j_A \right) J_A \right) J M_J \right\rangle,
\end{align}
and for the outgoing channels, it is given by
\begin{align}
\left| r_x r_b \alpha_{\text{out}} M_J \right\rangle = \left| r_x r_b \left( \left( l_x \left( j_x j_A \right) s_{xA} \right) J_x \left( \lambda_b j_b \right) J_b \right) J M_J \right\rangle,
\end{align}
where $j_b$, $j_x$, and $j_A$ denote the internal spins of particles $b$, $x$, and $A$, respectively; $s_{bx}$ and $s_{xA}$ represent the total spins of the $b$-$x$ and $x$-$A$ subsystems in the incoming and outgoing channels, respectively; $l_a$, $\lambda_a$, $l_x$, and $\lambda_b$ are the relative orbital angular momenta of the $b$-$x$, $a$-$A$, $x$-$A$, and $b$-$B^*$ pairs, respectively; and $J_a$ ($J_A$) and $J_x$ ($J_b$) are the total angular momenta of the subsystem (and spectator) in the incoming and outgoing channels, respectively.

The angular momentum basis can be normalized as follows:
\begin{align}
\left\langle r_{bx}' r_a' \alpha_{\text{in}}' \mid r_{bx} r_a \alpha_{\text{in}} \right\rangle = \frac{\delta\left( r_{bx}' - r_{bx} \right)}{r_{bx}' r_{bx}} \frac{\delta\left( r_a' - r_a \right)}{r_a' r_a} \delta_{\alpha_{\text{in}}', \alpha_{\text{in}}},
\end{align}
with a similar normalization applying to the outgoing basis. Additionally, a two-body angular momentum basis for the $x$-$A$ subsystem is defined as
\begin{align}
\left| r_x \beta M_x \right\rangle = \left| r_x \left( l_x s_{xA} \right) J_x M_x \right\rangle.
\end{align}

Consequently, the three-body outgoing state can be decoupled into a product of subsystem states:
\begin{align}
\left| r_x r_b \alpha_{\text{out}} M_J \right\rangle = \sum_{M_x M_b} \left\langle J_x M_x J_b M_b \mid J M_J \right\rangle \left| r_x \beta M_x \right\rangle \left| r_b J_b M_b \right\rangle,
\end{align}
and similarly, the incoming state can be written as
\begin{align}
\left| r_{bx} r_a \alpha_{\text{in}} M_J \right\rangle = \sum_{M_a M_A} \left\langle J_a M_a J_A M_A \mid J M_J \right\rangle \left| r_{bx} J_a M_a \right\rangle \left| r_a J_A M_A \right\rangle,
\end{align}
where $M_x$, $M_b$, $M_a$, and $M_A$ are the third components of $J_x$, $J_b$, $J_a$, and $J_A$, respectively.

The source term in the IAV model, within the zero-range approximation in the partial-wave basis, can be expressed as:
\begin{align}
\begin{aligned}
\langle r_x \beta M_x m_{j_b} \mid \rho M_a m_{j_A} \rangle 
&= \langle r_x \beta M_x m_{j_b} \chi_b^{(-)} \mid V_{bx} \mid \chi_a^{(+)} \phi_a M_a m_{j_A} \rangle \\
&= \sum_{\alpha_{\text{out}}, \alpha_{\text{in}}} \sum_{M_J} \sum_{M_b} \sum_{M_A} \int r_b^2 r_a^2 r_{bx}^2 \, dr_a \, dr_b \, dr_{bx} \, \langle J_x M_x J_b M_b \mid J M_J \rangle \langle J_a M_a J_A M_A \mid J M_J \rangle \delta_{\beta, \alpha_{\text{out}}} \\
&\quad \times \langle \chi_b^{(-)} m_{j_b} \mid r_b J_b M_b \rangle \int \mathcal{G}_{\alpha_{\text{in}}, \alpha_{\text{out}}}^{\text{out} \gets \text{in}} (\mathbf{r}_{bx}, \mathbf{r}_x) V_{bx}(r_{bx}) \langle r_a J_A M_A \mid \chi_a^{(+)} m_{j_A} \rangle \langle r_{bx} J_a M_a \mid \phi_a M_a \rangle \\
&\quad \times \frac{\delta(\mathbf{r}_b - \mathbf{f}(\mathbf{r}_{bx}, \mathbf{r}_x))}{r_b^2} \frac{\delta(\mathbf{r}_a - \mathbf{g}(\mathbf{r}_{bx}, \mathbf{r}_x))}{r_a^2} \, d\Omega_{r_{bx}} \, d\Omega_{r_x},
\end{aligned}
\end{align}
where
\begin{align}
\langle \chi_b^{(-)} m_{j_b} \mid r_b J_b M_b \rangle = \frac{4\pi}{k_b r_b} i^{-\lambda_b} e^{i\sigma_{\lambda_b}} f_{\lambda_b}(k_b r_b) \sum_{m_{\lambda_b}} \langle \lambda_b m_{\lambda_b} j_b m_{j_b} \mid J_b M_b \rangle Y_{\lambda_b}^{m_{\lambda_b}}(\hat{k}_b),
\end{align}
\begin{align}
\langle r_a J_A M_A \mid \chi_a^{(+)} m_{j_A} \rangle = \frac{4\pi}{k_a r_a} i^{\lambda_a} e^{i\sigma_{\lambda_a}} f_{\lambda_a}(k_a r_a) \sum_{m_{\lambda_a}} \langle \lambda_a m_{\lambda_a} j_A m_{j_A} \mid J_A M_A \rangle Y_{\lambda_a}^{m_{\lambda_a} *}(\hat{k}_a),
\end{align}
\begin{align}
\langle r_{bx} J_a M_a \mid \phi_a M_a \rangle = \frac{u_{l_a}(r_{bx})}{r_{bx}},
\end{align}
and
\begin{align}
\begin{aligned}
\mathcal{G}_{\alpha_{\text{in}}, \alpha_{\text{out}}}^{\text{out} \gets \text{in}} (\mathbf{r}_{bx}, \mathbf{r}_x) &= \sum_{L S} (2 S + 1) \sqrt{(2 J_a + 1)(2 J_A + 1)(2 J_x + 1)(2 J_b + 1)} \begin{Bmatrix} l_x & s_{xA} & J_x \\ \lambda_b & j_b & J_b \\ L & S & J \end{Bmatrix} \begin{Bmatrix} l_a & s_{bx} & J_a \\ \lambda_a & j_A & J_A \\ L & S & J \end{Bmatrix} \\
&\quad \times (-)^{s_{bx} + 2 j_A + j_x + j_b} \sqrt{(2 s_{xA} + 1)(2 s_{bx} + 1)} \begin{Bmatrix} j_A & j_x & s_{xA} \\ j_b & S & s_{bx} \end{Bmatrix} \sum_{M = -L}^{L} \sum_{m'_{l_x} m'_{\lambda_b}} \sum_{m'_{l_a} m'_{\lambda_a}} \\
&\quad \times \langle l_x m'_{l_x} \lambda_b m'_{\lambda_b} \mid L M \rangle \langle l_a m'_{l_a} \lambda_a m'_{\lambda_a} \mid L M \rangle Y_{l_x}^{m'_{l_x} *}(\hat{r}_x) Y_{\lambda_b}^{m'_{\lambda_b} *}(\hat{f}) Y_{l_a}^{m'_{l_a}}(\hat{r}_{bx}) Y_{\lambda_a}^{m'_{\lambda_a}}(\hat{g}).
\end{aligned}
\end{align}

Here, we assume $\hat{r}_{bx}$ is aligned with the $z$-direction, so $\int d\Omega_{r_{bx}} = 4\pi$. In the zero-range approximation, we set $\hat{r}_x = \hat{r}_a = \hat{r}_b$, $r_b = c r_x$, and $r_a = r_x$, where $c$ is the mass ratio of $A$ to $B^*$. Also, we use the known relations:
\begin{align}
Y_{l_a}^{m'_{l_a}}(\hat{r}_{bx}) = \left( \frac{2 l_a + 1}{4\pi} \right)^{1/2} \delta_{m'_{l_a}, 0},
\end{align}
\begin{align}
Y_l^{m *}(\theta, \varphi) = (-)^m Y_{l, -m}(\theta, \varphi),
\end{align}
\begin{align}
\begin{aligned}
\int Y_{l_x}^{m'_{l_x} *}(\hat{r}_x) Y_{\lambda_b}^{m'_{\lambda_b} *}(\hat{r}_x) Y_{\lambda_a}^{m'_{\lambda_a}}(\hat{r}_x) \, d\Omega_{r_x} &= (-)^{l_x + \lambda_b - \lambda_a} \left[ \frac{(2 l_x + 1)(2 \lambda_b + 1)}{4\pi (2 \lambda_a + 1)} \right]^{1/2} \\
&\quad \times \langle l_x 0 \lambda_b 0 \mid \lambda_a 0 \rangle \langle l_x m'_{l_x} \lambda_b m'_{\lambda_b} \mid \lambda_a m'_{\lambda_a} \rangle.
\end{aligned}
\end{align}

Thus, the source term can be rewritten as
\begin{align}
\begin{aligned}
\langle r_x \beta M_x m_{j_b} \mid \rho M_a m_{j_A} \rangle &= \frac{16\pi^2}{k_a k_b} \sum_{\alpha_{\text{out}}, \alpha_{\text{in}}} \sum_{M_J} \sum_{M_A} \sum_{M_b} \sum_{m_{\lambda_b}} \sum_{m_{\lambda_a}} \delta_{\beta, \alpha_{\text{out}}} \langle J_x M_x J_b M_b \mid J M_J \rangle \langle J_a M_a J_A M_A \mid J M_J \rangle \\
&\quad \times \langle \lambda_b m_{\lambda_b} j_b m_{j_b} \mid J_b M_b \rangle \langle \lambda_a m_{\lambda_a} j_A m_{j_A} \mid J_A M_A \rangle Y_{\lambda_a}^{m_{\lambda_a} *}(\hat{k}_a) Y_{\lambda_b}^{m_{\lambda_b}}(\hat{k}_b) \Lambda(r_x, \alpha_{\text{in}}, \alpha_{\text{out}}),
\end{aligned}
\end{align}
where
\begin{align}
\label{eq:lamba}
\Lambda(r_x, \alpha_{\text{in}}, \alpha_{\text{out}}) = i^{\lambda_a - \lambda_b} e^{i(\sigma_{\lambda_a} + \sigma_{\lambda_b})} \frac{f_{\lambda_b}(k_b c r_x)}{c r_x} \frac{f_{\lambda_a}(k_a r_x)}{r_x} \mathfrak{G}_{\alpha_{\text{in}}, \alpha_{\text{out}}} D_0,
\end{align}
with
\begin{align}
D_0 = (4\pi)^{1/2} \int r_{bx}^2 V_{bx}(r_{bx}) \langle r_{bx} J_a M_a \mid \phi_a M_a \rangle \, dr_{bx},
\end{align}
\begin{align}
\begin{aligned}
\mathfrak{G}_{\alpha_{\text{in}}, \alpha_{\text{out}}} &= \left[ \frac{(2 l_x + 1)(2 \lambda_b + 1)(2 l_a + 1)(2 J_a + 1)(2 J_A + 1)(2 J_x + 1)(2 J_b + 1)}{4\pi (2 \lambda_a + 1)} \right]^{1/2} (-1)^{l_x + \lambda_b - \lambda_a} \langle l_x 0 \lambda_b 0 \mid \lambda_a 0 \rangle \\
&\quad \times \sum_{L S} (2 S + 1) \begin{Bmatrix} l_x & s_{xA} & J_x \\ \lambda_b & j_b & J_b \\ L & S & J \end{Bmatrix} \begin{Bmatrix} l_a & s_{bx} & J_a \\ \lambda_a & j_A & J_A \\ L & S & J \end{Bmatrix} \\
&\quad \times (-)^{s_{bx} + 2 j_A + j_x + j_b} \sqrt{(2 s_{xA} + 1)(2 s_{bx} + 1)} \begin{Bmatrix} j_A & j_x & s_{xA} \\ j_b & S & s_{bx} \end{Bmatrix} \sum_{m'_{\lambda_a}} \langle l_a 0 \lambda_a m'_{\lambda_a} \mid L m'_{\lambda_a} \rangle \delta_{\lambda_a, L}.
\end{aligned}
\end{align}

It should be noted that in the zero-range approximation, the condition $\delta_{\lambda_a, L}$ is required. Thus, only the $l_a = 0$ zero-range case provides a good approximation to the finite-range results. For cases where $l_a \neq 0$, the zero-range source term yields zero, even when the finite-range source term is significant.

Then, the $\varphi_x$ function in the partial wave basis with the zero-range approximation becomes:
\begin{align}
\begin{aligned}
\langle r_x \beta M_x m_{j_b} \mid \varphi_x(\mathbf{k}_b) M_a m_{j_A} \rangle &= \sum_{\alpha_{\text{out}}, \alpha_{\text{in}}} \sum_{M_J} \sum_{M_A} \sum_{M_b} \sum_{m_{\lambda_b}} \sum_{m_{\lambda_a}} \delta_{\beta, \alpha_{\text{out}}} \langle J_x M_x J_b M_b \mid J M_J \rangle \langle J_a M_a J_A M_A \mid J M_J \rangle \\
&\quad \times \langle \lambda_b m_{\lambda_b} j_b m_{j_b} \mid J_b M_b \rangle \langle \lambda_a m_{\lambda_a} j_A m_{j_A} \mid J_A M_A \rangle Y_{\lambda_a}^{m_{\lambda_a} *}(\hat{k}_a) Y_{\lambda_b}^{m_{\lambda_b}}(\hat{k}_b) \mathcal{R}(r_x, \alpha_{\text{in}}, \alpha_{\text{out}}),
\end{aligned}
\end{align}
where
\begin{align}
\mathcal{R}(r_x, \alpha_{\text{in}}, \alpha_{\text{out}}) = -\frac{32 \pi^2 \mu_x}{\hbar^2 k_a k_b k_x} \frac{1}{c r_x} i^{\lambda_a - \lambda_b} e^{i(\sigma_{\lambda_a} + \sigma_{\lambda_b})} \mathfrak{G}_{\alpha_{\text{in}}, \alpha_{\text{out}}} D_0 \Pi(r_x),
\end{align}
with
\begin{align}
\Pi(r_x) = \int \frac{f_{\beta}(r_x' <) h_{\beta}^{(+)}(r_x' >)}{r_x'} f_{\lambda_b}(k_b c r_x') f_{\lambda_a}(k_a r_x') \, dr_x'.
\end{align}

In the Vincent-Fortune method, under the zero-range approximation, we only consider the external part. For a given $r_x$,
\begin{align}
\Pi(r_x) = \Pi^{\text{int}}(r_x) + \Pi^{\text{ext}}(r_x).
\end{align}
In the current approach, I use the finite-range method to compute $\Pi^{\text{int}}$ and the zero-range method to compute $\Pi^{\text{ext}}$. For the asymptotic behavior, one has:
\begin{align}
f_{\lambda_b}(k_b c r_x) &= \frac{i}{2} \left[ H_{\lambda_b}^{(-)}(k_b c r_x) - S_{\lambda_b} H_{\lambda_b}^{(+)}(k_b c r_x) \right], \\
f_{\lambda_a}(k_a r_x) &= \frac{i}{2} \left[ H_{\lambda_a}^{(-)}(k_a r_x) - S_{\lambda_a} H_{\lambda_a}^{(+)}(k_a r_x) \right],
\end{align}
\begin{align}
\begin{aligned}
\Pi^{\text{ext}}(r_x) &= -\frac{1}{4} \int_{R}^\infty \frac{dr_x'}{r_x'} f_{\beta}(r_x) \Big[ H_{l_x}^{(+)}(k_x r_x') H_{\lambda_b}^{(-)}(k_b c r_x') H_{\lambda_a}^{(-)}(k_a r_x') - S_{\lambda_b} H_{l_x}^{(+)}(k_x r_x') H_{\lambda_b}^{(+)}(k_b c r_x') H_{\lambda_a}^{(-)}(k_a r_x') \\
&\quad - S_{\lambda_a} H_{l_x}^{(+)}(k_x r_x') H_{\lambda_b}^{(-)}(k_b c r_x') H_{\lambda_a}^{(+)}(k_a r_x') + S_{\lambda_a} S_{\lambda_b} H_{l_x}^{(+)}(k_x r_x') H_{\lambda_b}^{(+)}(k_b c r_x') H_{\lambda_a}^{(+)}(k_a r_x') \Big].
\end{aligned}
\end{align}

We treat the above equation in the complex plane to overcome the divergence of the integration on the real axis. For example, we consider the first product term:
\begin{align}
\begin{aligned}
&H_{l_x}^{(+)}(k_x z) H_{\lambda_b}^{(-)}(k_b c z) H_{\lambda_a}^{(-)}(k_a z) \\
&\xrightarrow{z \to \infty} e^{i k_x z} e^{-i k_b c z} e^{-i k_a z} \\
&= e^{i (k_x - c k_b - k_a) (x + i y)} \\
&= e^{i (k_x - c k_b - k_a) x} e^{- (k_x - c k_b - k_a) y}.
\end{aligned}
\end{align}

Now, if $-(k_x - c k_b - k_a) > 0$, the second exponential will diverge unless $y$ takes a negative value. Thus, in the complex analysis, one should evaluate this term in the lower half of the complex plane.

Proceeding in a similar manner with the other terms, we have two groups of product terms, $L(z)$ and $U(z)$, where one is analytic in the lower half and the other in the upper half of the complex plane:
\begin{align}
\begin{aligned}
\label{eq:integrand}
L(z) &= \frac{1}{z} \left[ H_{l_x}^{(+)}(k_x z) H_{\lambda_b}^{(-)}(k_b c z) H_{\lambda_a}^{(-)}(k_a z) - S_{\lambda_b} H_{l_x}^{(+)}(k_x z) H_{\lambda_b}^{(+)}(k_b c z) H_{\lambda_a}^{(-)}(k_a z) \right], \\
U(z) &= \frac{1}{z} \left[ -S_{\lambda_a} H_{l_x}^{(+)}(k_x z) H_{\lambda_b}^{(-)}(k_b c z) H_{\lambda_a}^{(+)}(k_a z) + S_{\lambda_a} S_{\lambda_b} H_{l_x}^{(+)}(k_x z) H_{\lambda_b}^{(+)}(k_b c z) H_{\lambda_a}^{(+)}(k_a z) \right].
\end{aligned}
\end{align}

Moreover, as $|z| \to \infty$, $L(z) \to 0$ and $U(z) \to 0$. We can define three integration paths: $C_1$ is the real axis, $C_2$ is in the upper plane, and $C_3$ is in the lower plane:
\begin{align}
\begin{aligned}
\Pi^{\text{ext}}(r_x) &= -\frac{f_{\beta}(r_x)}{4} \int_{C_1} dz \, L(z) - \frac{f_{\beta}(r_x)}{4} \int_{C_1} dz \, U(z) \\
&= -\frac{f_{\beta}(r_x)}{4} \int_{C_2} dz \, U(z) - \frac{f_{\beta}(r_x)}{4} \int_{C_2'} dz \, U(z) - \frac{f_{\beta}(r_x)}{4} \int_{C_3} dz \, L(z) - \frac{f_{\beta}(r_x)}{4} \int_{C_3'} dz \, L(z) \\
&= -\frac{f_{\beta}(r_x)}{4} \int_{C_2} dz \, U(z) - \frac{f_{\beta}(r_x)}{4} \int_{C_3} dz \, L(z).
\end{aligned}
\end{align}

Here, we used the fact that, if the radius of the semicircle is taken to be large, then $\int_{C_2'} dz \, U(z) = 0$ and $\int_{C_3'} dz \, L(z) = 0$.

Now,
\begin{align}
\begin{array}{ll}
C_2 := R + i y, & 0 \leq y < \infty, \\
C_3 := R - i y, & 0 \leq y < \infty,
\end{array}
\end{align}

Therefore, one has:
\begin{align}
\Pi^{\text{ext}}(r_x) = -\frac{f_{\beta}(r_x)}{4} \left[ i \int_0^\infty U(R + i y) \, dy - i \int_0^\infty L(R - i y) \, dy \right],
\end{align}
with $U(R_{\max} + i y) \to 0$ and $L(R_{\max} - i y) \to 0$ as $y \to \infty$.

\end{document}